\documentclass{aa}

\usepackage{graphicx}
\usepackage{txfonts}
\usepackage{natbib}

\catcode`\@=11
\def\gsim{\ifmmode{\mathrel{\mathpalette\@versim>}}
    \else{$\mathrel{\mathpalette\@versim>}$}\fi}
\def\lsim{\ifmmode{\mathrel{\mathpalette\@versim<}}
    \else{$\mathrel{\mathpalette\@versim<}$}\fi}
\def\@versim#1#2{\lower 2.9truept \vbox{\baselineskip 0pt \lineskip
    0.5truept \ialign{$\m@th#1\hfil##\hfil$\crcr#2\crcr\sim\crcr}}}
\catcode`\@=12

\begin{document}
\input{psfig.sty}

\title{The Metal Content of Bulge Field Stars from FLAMES-GIRAFFE Spectra. I. 
Stellar parameters and Iron Abundances
\thanks{Observations collected at the European Southern Observatory, Paranal, 
Chile (ESO programmes 71.B-0617 and 73.B-0074). }}
\author{
M. Zoccali\inst{1}
\and
V. Hill\inst{2}
\and
A. Lecureur\inst{2,3}
\and
B. Barbuy\inst{4}
\and
A. Renzini\inst{5}
\and
D. Minniti\inst{1}
\and
A. G\'omez\inst{2}
\and
S. Ortolani\inst{6}
\fnmsep
}
\offprints{M. Zoccali}
\institute{
P. Universidad Cat\'olica de Chile, Departamento de Astronom\'\i a y 
Astrof\'\i sica, Casilla 306, Santiago 22, Chile;\\
\email{mzoccali@astro.puc.cl, dante@astro.puc.cl}
\and  
GEPI, Observatoire de Paris, CNRS, Universit\'e Paris Diderot; Place Jules Janssen 92190
Meudon, France;\\
\email{Vanessa.Hill@obspm.fr, Aurelie.Lecureur@obspm.fr, Ana.Gomez@obspm.fr}
\and
Astronomisches Rechen-Institut, Zentrum f\"ur Astronomie der Universit\"at Heidelberg,
M\"onchhofstr. 12-14, D-69120 Heidelberg, Germany;
\and
Universidade de S\~ao Paulo, IAG, Rua do Mat\~ao 1226,
Cidade Universit\'aria, S\~ao Paulo 05508-900, Brazil;\\
\email{barbuy@astro.iag.usp.br}
\and
INAF - Osservatorio Astronomico di Padova, Vicolo dell'Osservatorio 2,
I-35122 Padova, Italy; \\
\email{alvio.renzini@oapd.inaf.it}
\and
Universit\`a di Padova, Dipartimento di Astronomia, Vicolo
dell'Osservatorio 5, I-35122 Padova, Italy;\\
\email{sergio.ortolani@unipd.it}
}

\date{Received ; accepted }
                                                              
\abstract
{}
{We determine the iron   distribution function (IDF) for  bulge  field
stars, in three different fields along  the Galactic minor axis and at
latitudes $b=-4^\circ$, $b=-6^\circ$,   and $b=-12^\circ$.   A  fourth
field including NGC6553 is also included in the discussion.}
{About 800  bulge field  K  giants   were observed  with the   GIRAFFE
spectrograph    of     FLAMES@VLT    at       spectral      resolution
R$\sim$20,000.    Several of them were observed    again  with UVES at
R$\sim$45,000  to insure the  accuracy   of the measurements. The  LTE
abundance analysis yielded stellar parameters and iron abundances that
allowed us to construct an IDF for the bulge that, for the first time,
is based on high-resolution spectroscopy for each individual star.}
{The IDF  derived   here is  centered on solar  metallicity,  and
extends  from [Fe/H]$\sim$-1.5  to [Fe/H]$\sim$+0.5.  The distribution
is asymmetric, with a sharper cutoff on the high-metallicity side, and
it  is narrower  than previously  measured.   A variation in  the mean
metallicity along the bulge minor axis is clearly between $b=-4^\circ$
and  $b=-6^\circ$ ([Fe/H] decreasing $\sim $by  0.6 dex per kpc).  The
field at $b=-12^\circ$ is consistent with the  presence of a gradient,
but   its quantification  is   complicated  by the  higher  disk/bulge
fraction in this field.}
{Our findings support a scenario in which both infall and outflow were
important during the bulge formation, and then suggest the presence of a
radial gradient, which poses some challenges to the scenario in which
the bulge would result solely from the vertical heating of the bar.}

\titlerunning{Galactic Bulge Iron abundance}
\authorrunning{M. Zoccali et al.}

\maketitle

\section{Introduction}

The  Galactic bulge is  the  nearest  galactic  spheroid, and it can be
studied in  greater  detail than  any other  one.  In  particular, its
stellar content  can be characterized  in terms of age and composition
distribution functions,  coupled  with  kinematical information. Thus,
the bulge offers a unique opportunity to  construct the star formation
and   mass assembly history of  a   galactic spheroid, hence providing
a unique benchmark for theories of  galaxy formation. The Galactic bulge
is dominated by stellar populations older than $\sim 10$ Gyr (Ortolani
et al.  1995; Feltzing \& Gilmore 2000; Zoccali  et al.  2003), with no
detectable trace of younger  stellar populations. As a result, most of its
stars were formed at a cosmic epoch that  corresponds to $z\gsim 2$, making
its   study  quite complementary  to  that   of galaxies at  such high
redshifts.

Starting with the pioneering spectroscopic  study of Rich (1988),  the
distribution function of the iron abundance among bulge stars has been
further explored  and refined by McWilliam  \& Rich (1994), Ibata
\&  Gilmore  (1995a,b)  Minniti    (1996), Sadler   et  al.   (1996),
Ram\'{\i}rez et al.  (2000),  and Fulbright, McWilliam \& Rich (2006),
using spectroscopic observations, and by Zoccali et al.  (2003) with a
purely photometric method.  Among them,  the McWilliam \& Rich  (1994)
and Fulbright et al.   (2006) analyses deserve special mention because
they  were the only  ones to  obtain high-resolution spectra, although
only for a small  sample of stars  (11 and 27, respectively), used  to
calibrate some previous, low-resolution  analysis of a larger  sample.
The  choice   of  this method     was   dictated by  high   resolution
spectroscopic surveys being  carried out with long-slit spectrographs,
thus observing  just one or two stars  at a time.   With the advent of
the FLAMES multiobject spectroscopic facility at  the VLT (Pasquini et
al.  2003)   it then became  possible  to  observe a   large number of
objects simultaneously, at high spectral resolution, a quantum jump in
this kind of studies.

FLAMES feeds 8 fibers to   the UVES high resolution spectrograph,  and
over   130   fibres   to    the   GIRAFFE  medium-high      resolution
spectrograph. The results  of bulge stars  observations of 50 K giants
obtained with UVES with $R\simeq 45,000$ have been reported by Zoccali
et    al.   (2006) concerning the    oxygen  abundance and  the [O/Fe]
vs. [Fe/H] correlation,  and by Lecureur  et al. (2007) concerning the
abundance of O, Na, Mg, and Al.

This paper is the first of  a series devoted  to the detailed chemical
analysis  of  a sample  of 720 bulge  giant stars,   in four different
fields,      observed    with  FLAMES-GIRAFFE     with  a   resolution
R$\sim$20,000. Another 220 bulge red clump stars were observed, in the
same condition, as  part of the GIRAFFE  GTO programme (Lecureur et al.
2008).  The  latter sample is   occasionally combined with the present
one, in order to investigate some of  the systematics and increase the
statistics.     Taking  advantage of  the    FLAMES  link to the  UVES
spectrograph, 58  target  stars  were {\it  also}  observed  at higher
spectral    resolution    ($R=45,000$),  making it possible  to  compare
abundances derived from medium and high resolution spectra.

\section{Observations}

\begin{table}[ht!]
\caption{Characteristics of the four bulge fields.}
\label{fields}
\begin{tabular*}{0.47\textwidth}{@{\extracolsep{-2pt}}llccrcc}
\hline\hline
\noalign{\smallskip}
Nr.  &  Identification       &  $l$ & $b$ & R$_{\rm GC}$ & E$(B-V)$ &  N$_{\rm stars}$  \\
     &                       &      &     &   (pc)       &          &             \\
\hline
   1  & Baade's Window       &  1.14 & $-4.18$ &  604 & 0.55 & 204  \\
   2  & $b=-6^\circ$ Field   &  0.21 & $-6.02$ &  850 & 0.48 & 213  \\
   3  & $b=-12^\circ$ Field  &  0.00 & $-12.0$ & 1663 & 0.20 & 104  \\
   4  & NGC~6553 Field       &  5.25 & $-3.02$ &  844 & 0.70 & 201 \\
\hline
\end{tabular*}
\end{table}

Spectra for  a sample  of  K giants   in four bulge  fields  have been
collected  at the VLT-UT2   with  the FLAMES-GIRAFFE  spectrograph, at
resolution R$\sim$20,000. A  total wavelength range of $\sim$760 $\AA$
has been   covered  through  the  setup  combinations   HR13+HR14+HR15
(programme  071.B-0617) for fields~1  and 2 in Table~\ref{fields}, and
HR11+HR13+HR15  (programme   073.B-0074) for   fields~3  and 4.    The
characteristics of the observed fields,   together with the number  of
target stars contained in each, are listed in Table~\ref{fields}.  The
total exposure time  varies  from  about 1 hour   to almost  5  hours,
depending on the  setup and on the  star luminosity (targets have been
divided into a bright and  a faint group)  in order to insure that the
final S/N of  each coadded spectrum is $\sim  60$. In fact, the actual
S/N  is  not   identical among the    targets of  a given  field  (see
Table~\ref{targbox}) due to the differences  both in magnitude and  in
the average accuracy of fibre positioning.

\subsection{Photometric Data}

\begin{table}[ht!]
\caption{Magnitude, color and S/N range of the spectroscopic targets.}
\label{targbox}
\begin{tabular*}{0.47\textwidth}{@{\extracolsep{-2pt}}llccc}
\hline\hline
\noalign{\smallskip}
Nr.  &  Identification       &  $(V-I)$ &  $I$   & typical S/N \\
     &                       &  range   &  range & @6200 $\AA$ \\
\hline
   1  & Baade's Window       &  1.53$-$2.62 & 14.20$-$14.70 & 40$-$60 \\
   2  & $b=-6^\circ$ Field   &  1.41$-$2.43 & 14.00$-$14.50 & 60$-$90 \\
   3  & $b=-12^\circ$ Field  &  1.10$-$1.70 & 13.70$-$14.93 & 40$-$80 \\
   4  & NGC~6553 Field       &  2.00$-$3.00 & 14.00$-$14.50 & 30$-$60 \\
\hline
\end{tabular*}
\end{table}

\begin{figure*}[t]
\psfig{file=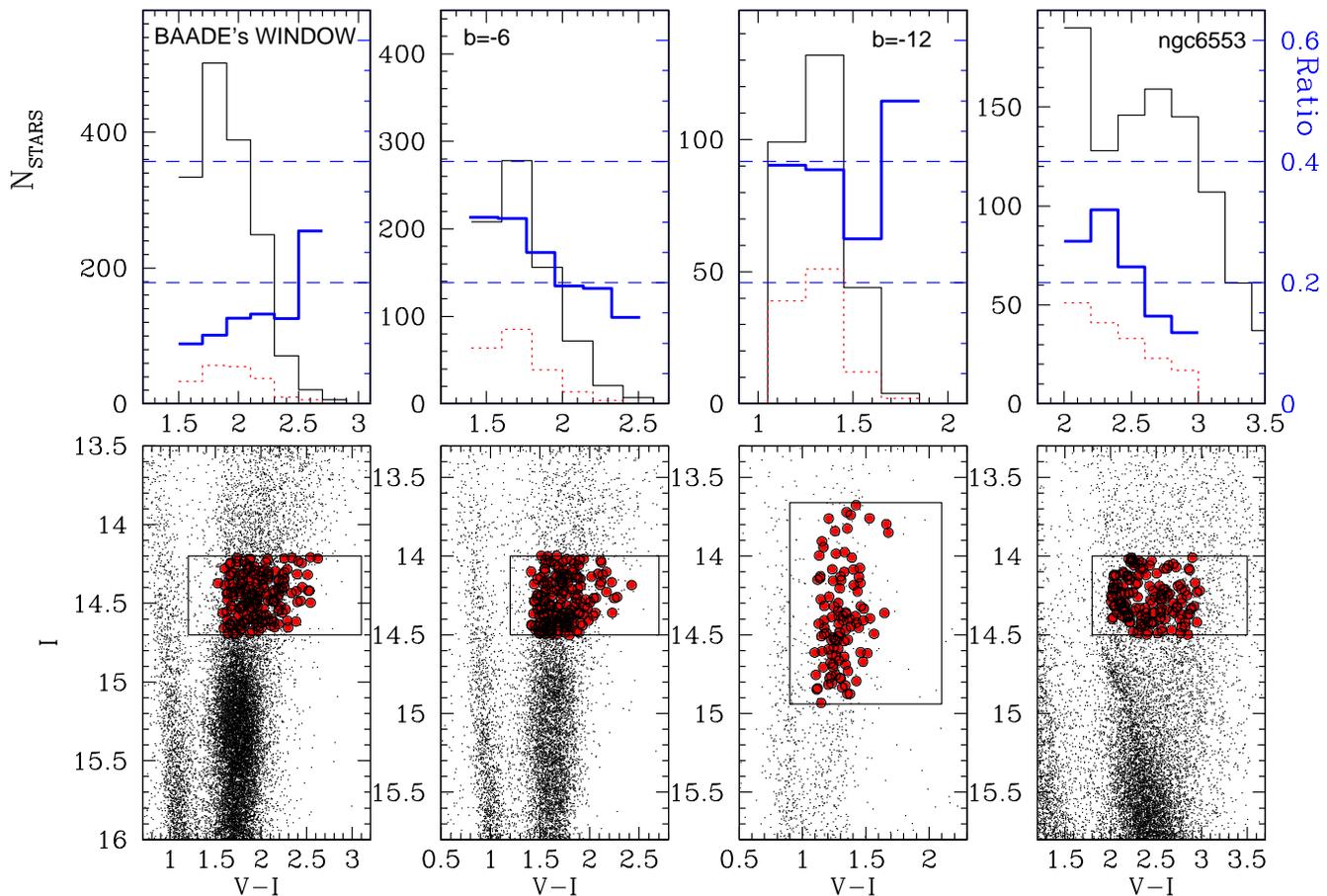,angle=-90,width=18cm}
\caption{Lower panels: the color magnitude diagram of the four observed
fields, with the spectroscopic target stars marked as large filled 
circles. From left to right the fields are: Baade's Window, the $b=-6^\circ$
field, the $b=-12^\circ$ field and the NGC 6553 field. The large 
box shows the magnitude limits used for target selection, and the color
limits of the upper panels. Upper panels: color histogram of the available
stars in the selected target box (thin solid), of the actually observed
targets (dotted) and of the ratio of observed to available stars (thick
solid). The vertical scale of the latter histogram is shown on the right
hand side of the upper panels, and two dashed lines have been drawn 
at 0.2 and 0.4 to help the eye in reading the figure. }
\label{cmds}
\end{figure*}

In the  color  magnitude diagram, these  stars are  located on the red
giant branch  (RGB),  roughly 1  magnitude  above the  red clump  (see
Table~\ref{targbox}), as shown in the lower panels of Fig.~\ref{cmds}.
The astrometry  and the   photometric $V,I$  data come  from  the OGLE
catalogue  (Udalski et al.   2002) for  our Field-1,  from archive WFI
images obtained within the ESO Pre-FLAMES survey  (e.g., Momany et al.
2001) from which our group obtained the stellar catalogue (Field-2
and -3), and from  archive WFI images  from proposal 69.D-0582A kindly
reduced by Yazan  Momany, for Field-4.  Cross-identification with  the
2MASS point source  catalogue (Carpenter  et  al. 2001) allowed us  to
obtain $V,I,J,H,K$ magnitudes  for each of the  target stars.  Some of
the  fields contain a globular  cluster, namely NGC6528 and NGC6522 in
Baade's Window, NGC6558 in the $b=-6^\circ$ field,  and NGC6553 in the
eponymous field.  Member stars of these  clusters will be discussed only
marginally here, because they are the subject of dedicated papers (see
Barbuy et al.  2007, for NGC6558).

In  order  to avoid  strong biases  in the resulting iron distribution
function (IDF)  we included targets spanning the  whole color range of
the RGB at that  magnitude.  However, the need  to maximize the number
of  targets while  avoiding   forbidden   fibre crossings,  makes   it
impossible to actually  fine tune a uniform sampling  of the RGB color
span.  The upper panels of  Fig.~\ref{cmds} show as a thick
histogram  the ratio of  the number of   targets (dotted histogram) at
each color  bin, to the available stars  (solid histogram).  The scale
of the  thick histogram  is shown on   the right side of  the  figure.
Since  we  expect   a correlation  between  the  RGB  color   and  the
metallicity of the stars, the ideal, unbiased sample would be one with
a   flat ratio   between  observed    and available   stars   at  each
color\footnote{Note  that a region within  2 arcmin from the cluster
center was removed both from the available-star and from the target
sample.}. As  stated above, it  is
virtually impossible to keep  this constraint all  the way through the
fibre   allocation process.   For  this reason,   further below in our
analysis we will correct the {\it raw} IDF for this  bias.  We will do
that in two independent ways, namely: {\it i)} we randomly remove from
the  IDF stars belonging to  the most  populated  color bins, until we
reach a flat   target/available-star ratio; or {\it  ii)}  in the less
populated color bins we duplicate some randomly-extracted stars, until
we reach a  flat target/available-star  ratio.  In both  cases we will
repeat  the    process 200 times,  in    order  to minimize stochastic
fluctuations in the final star list due to the random extractions, and
we  combine the   results just by  merging the   200  star lists.  The
resulting   IDF  from   the    two   methods  described    above   are
indistinguishable, proving that the method is indeed robust (see Fig.~\ref{mdfcorr} below.)

The $V-I$ color was used to obtain photometric temperatures, according
to the latest empirical calibration  (Ram\'{\i}rez \& Mel\'endez 2005)
based on  the InfraRed Flux Method.  As an additional indicator of the
star  temperature we measured  the strength  of  the TiO band using an
index defined  between  6190-6250  $\AA$ (band) and   6120-6155  $\AA$
(continuum region). The strength of this  index indeed correlates very
well with   the photometric temperatures,  for T$_{\rm  phot}<4500$~K,
where the TiO band is strong enough to be measured.  The TiO index was
used in two ways. First,  it allowed us  to  establish that the  $V-I$
color was the best one to  derive photometric temperatures, as the one
showing the  smallest dispersion  in the correlation   between T$_{\rm
V-I}$ and T$_{\rm TiO}$.  Second, since we expect that the $V-I$ color
is more sensitive  to differential reddening  variations than the  TiO
index, for stars cooler than 4500~K, we used  the latter to estimate a
$(V-I)_0$  color, to be  converted into a  photometric temperature.  The
calibration we used  to convert the  TiO index into a $(V-I)_0$  color
was estimated as a linear fit to the  observed correlation between the
strength of  the index and the  $(V-I)_0$ obtained assuming a constant
reddening for each  field (see Table~\ref{fields}). Therefore, the use
of the T$_{\rm TiO}$  for the coolest stars  only minimizes the effect
of the  {\it  differential} reddening.  Any   problem  related to  the
adopted  color-temperature calibration  by Ram\'{\i}rez \&  Mel\'endez
(2005) will obviously be present also in our TiO-$(V-I)_O$-temperature
calibration.   Finally, it is  worth  emphasizing that the photometric
temperature has  only been used as  an initial  first guess. The final
adopted  temperature  is the    spectroscopic one,  derived   imposing
excitation equilibrium on a sample of $\sim 60$ FeI lines.

Photometric gravity   was   instead calculated  from    the  classical
relation:   

\noindent
$ \log g_*= \log g_\odot+4 \log \frac{T_*}{T_\odot}+
0.4(M_{\rm bol}-M_{\rm bol, \odot})+\log \frac{M_*}{M_{\odot}} $
 
\noindent
adopting a mean distance of $8$ kpc for the bulge, T$_{\odot}$=5770 K,
$\log   g_{\odot}$=4.44,  M$_{\rm bol    \odot}$=4.75 and   M$_*$=0.85
M$_{\odot}$.  Note that,  at each step  of the iterative process  to
converge on the stellar parameters  and metallicity, described  below,
the photometric gravity was  re-calculated using the  appropriate (now
spectroscopic)  temperature and metallicity   (to compute the $V$-band
bolometric   correction, from Alonso et  al.  1999) for the star under
analysis.

\subsection{Spectroscopic Data}

Individual   spectra     were     reduced   with      the     GIRBLDRS
pipeline\footnote{Available               at               SouceForge,
http://girbldrs.sourceforge.net .}  provided by  the FLAMES consortium
(Geneva Observatory; Blecha et al. 2003), including bias, flatfield,
extraction  and wavelength calibration. All the  spectra for each star
(a number between    1 and 5,    depending  on the  field) were   then
registered in wavelength  to correct for  heliocentric radial velocity
and coadded to a single spectrum per setup,  per star.  In each plate,
about 20 GIRAFFE  fibres were allocated to empty  sky  regions.  These
sky spectra were visually inspected to  reject the few that might have
evident stellar flux, and then coadded to a single sky spectrum.  The
latter was then subtracted from the spectrum of each target star.  The
equivalent widths (EWs)  for selected iron  lines were measured  using
the automatic code DAOSPEC (Stetson \& Pancino, in preparation\footnote{
http://cadcwww.hia.nrc.ca/stetson/daospec/}.)

\section{Line list}

The selection of a clean line list, and the compilation of their atomic
parameters, has been done with special care.   An initial line list was
compiled from  NIST (Fuhr \& Wiese  2006). Each line  was then checked
against blends,  in the relevant   metallicity and temperature  range,
using synthetic spectra generated with and without the line, using the
codes by Alvarez \& Plez (1998) and Barbuy et al.  (2003). The
oscillator strengths log gf
of each of the  clean lines were  then  modified by requiring  that it
would give [Fe/H]=+0.30 on the  spectrum of $\mu$ Leonis, observed  at
the Canada-France-Hawaii Telescope  with the ESPaDOnS spectrograph, at
resolution R=80,000 and S/N$\sim 500$.   The following parameters were
determined    for   $\mu$  Leo:   T$_{\rm   eff}$=4550   K,  logg=2.3,
microturbulence velocity $V_{\rm t}$=1.3  km/s.  The final line list is
thus the same used in Lecureur et al. (2007) with the addition of some
more lines in the region covered by  the HR11 GIRAFFE setup (5597-5728
\AA) not included in that paper.  With the same set of atomic lines we
obtained [Fe/H]=$-0.52$ for  Arcturus (T$_{\rm eff}$=4300~K,  logg=2.5
and $V_{\rm  t}$=1.5  km/s), and [Fe/H]=$-0.02$  for the  Sun (T$_{\rm
eff}$=5777~K,  logg=4.4   and $V_{\rm  t}$=1.0    km/s).  The  damping
constants, taken from   Coelho  et  al.  (2005)  were  computed  where
possible,  and  in particular for  most of   the  FeI lines, using the
collisional  broadening   theory   (Barklem,  Anstee  \&  O'Mara 1998;
Barklem, Piskunov \& O'Mara 2000).

\begin{figure}[ht]
\psfig{file=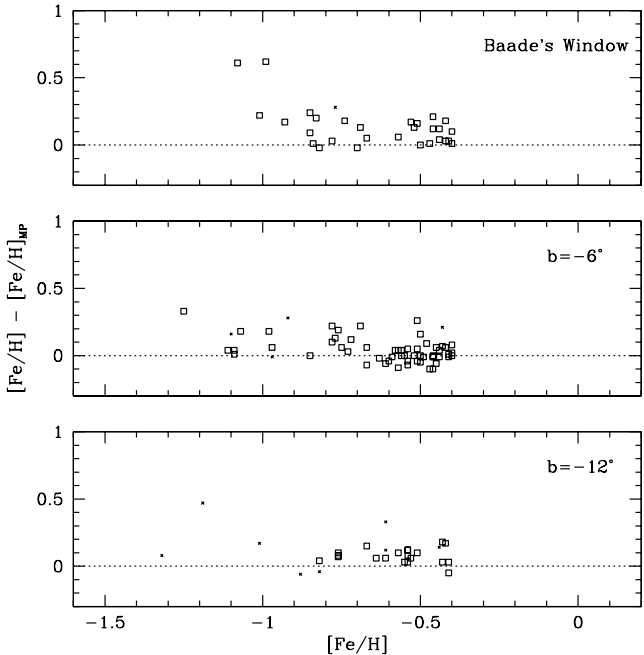,angle=0,width=9cm}
\caption{Comparison between the iron abundance obtained using either
the {\it metalpoor} or the {\it metalrich} line list, for stars 
with $[Fe/H]<-0.4$. Small symbols are stars for which the metal-rich 
line list was giving very poor results (e.g., it was not possible to
converge on excitation equilibrium, or the dispersion of measurements
from individual lines was too high) and those stars were thus marked
as ``low quality''.}
\label{mrmp}
\end{figure}

This line list proved effective  down  to [Fe/H]$\sim -0.8$,   including
lines with a wide range of EWs for all the stars. For more metal-poor
stars, however, we started lacking strong lines.  A different line list
was then compiled,  including lines that  would be too strong in $\mu$
Leo, but of intermediate strength in relatively metal poor stars. This
one was produced with the same  criterion described below, except that
we have kept the NIST log gfs. In order to  ensure a smooth transition
between the  so  called  {\it   metalrich} and   the {\it   metalpoor}
line list,  we   measured the  metallicity of   Arcturus,  from  a UVES
spectrum (Bagnulo et al. 2003) with  both, and switch  from one to the
other at  [Fe/H]=-0.4, where  we   check  that  the  two would    give
consistent results.     The metallicity  of  Arcturus with    the {\it
metalpoor} line list,  and the   same   parameters listed   above,   is
[Fe/H]=$-0.55$.

Figure~\ref{mrmp}  shows   the difference  between  the [Fe/H]  values
resulting from  the use of the {\it  metalrich} or the {\it metalpoor}
line list, for stars with  [Fe/H]$<-0.4$. It can  be  seen that at  the
transition region  ($-0.6<$[Fe/H]$<-0.4$) the  systematic  difference is
0.09 dex, 0.01 dex and 0.08  dex, for Baade's Window, the $b=-6^\circ$
field and the $b=-12^\circ$ field,  respectively. For more metal  poor
stars the difference obviously increases,  because the {\it metalrich}
line list is clearly not appropriate for them.

Finally, in order to complete the analysis of possible systematics due
to the adopted  line list, we measured the  metallicity  of Arcturus by
selecting only  the wavelength ranges  of the two setup combinations we
used for our  targets, namely  HR13+HR14+HR15 or HR11+HR13+HR15.   The
difference in [Fe/H] is +0.01  dex, the first setup combination giving
higher metallicity,  both with the {\it metalpoor}   and with the {\it
metalrich} line list. The value quoted above for the metallicity of
Arcturus ([Fe/H]=$-0.52$) refers to the HR13+HR14+HR15 setup combination.

\section{Abundance Analysis}

LTE  abundance analysis  was performed using   well tested procedures
(Spite 1967)  and the  new  MARCS spherical models (Gustafsson  et al.
2003;    available   at  http://www.marcs.astro.uu.se/).    Excitation
equilibrium  was  imposed  on  FeI  lines    in  order to  refine  the
photometric T$_{\rm eff}$,  while photometric gravity was imposed even
if  ionization equilibrium     was not fulfilled    (c.f.  Zoccali et
al.  2006).  The microturbulence  velocity ($V_{\rm t}$)  was found by
imposing a constant [Fe/H] for lines of different expected strengths (
predicted EWs for  a given stellar model).   The reason for the latter
choice is that, when  plotting  derived metallicities versus  observed
EWs, the obvious correlation of the errors (a too high EW would give a
too high [Fe/H], and  vice versa) would lead us  to detect a  positive
slope, hence  to increase the $V_{\rm t}$   (Magain 1984).  The effect
may be  negligible with very high  S/N, high resolution spectra (i.e.,
when the errors on the  EWs are also  negligible) but we verified that
it would introduce a significant  systematic error in the measurements
of  the present GIRAFFE spectra.  Extensive  discussion of this effect
can be found in Lecureur et al. (2008).

Finally, once converged  on the best  stellar parameters, we calculate
the  [Fe/H]  of each   star  as a  weighted  mean of  the line-by-line
measurements.  The weight  associated to  each  line is given  by  the
inverse square  of its abundance error,  as derived from  the error in
the measured EWs.

\section{Estimates of metallicity uncertainties}

In this Section we discuss all the available information about the
error associated to the [Fe/H] of each star. Some of our tests will
quantify only the statistical errors, some others will quantify a 
combination of part of the systematics and the statistical errors. 
Finally, we will try to combine all this information together in order
to estimate how far we can go in the interpretation of apparent features
of the obtained IDFs.

\subsection{Line-to-line dispersion}

\begin{figure}[hb!]
\psfig{file=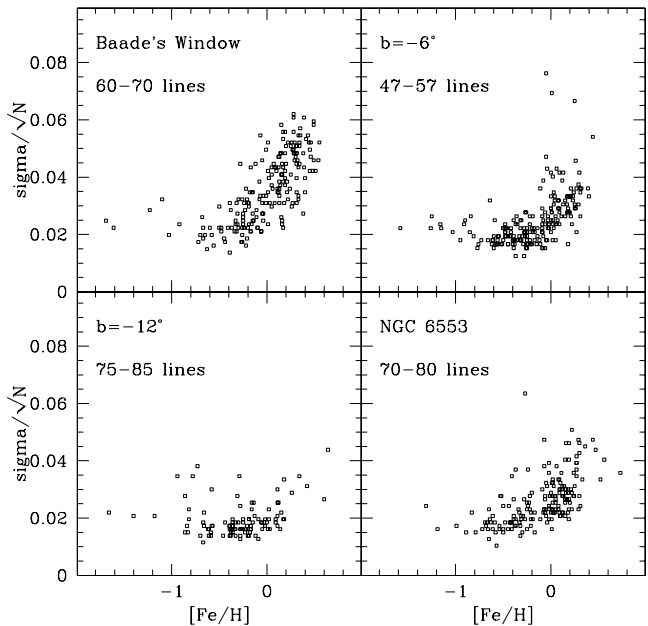,angle=0,width=9cm}
\caption{The error on the mean [Fe/H] of each star, due to the 
line-to-line dispersion. The minimum and maximum number of FeI lines
found in each star is indicated in the figure labels.}
\label{sigma}
\end{figure}

Figure~\ref{sigma}  shows   the scatter   in  the line-by-line  [Fe/H]
measurement, divided by the square root of the number of lines, versus
[Fe/H].  This  is  a  fairly reasonable  estimate  of  the statistical
(line-to-line    only)    fluctuation associated    with   each [Fe/H]
value. Clearly, the more metal rich the star,  the more crowded is the
spectrum, hence the  higher the dispersion  of  [Fe/H] from individual
lines.  Baade's Window's  stars show the largest  scatter, likely due to
the lower S/N  of those spectra, caused  by the lower accuracy of  the
astrometry (from  the OGLE catalogue) used  to position the fibres. In
any  case, the statistical error  from   the dispersion of  individual
lines is less than 0.06 dex for Baade's Window, and less than 0.04 dex
for the other fields.

\subsection{Degeneracy in the stellar parameters}

\begin{figure}[hb!]
\psfig{file=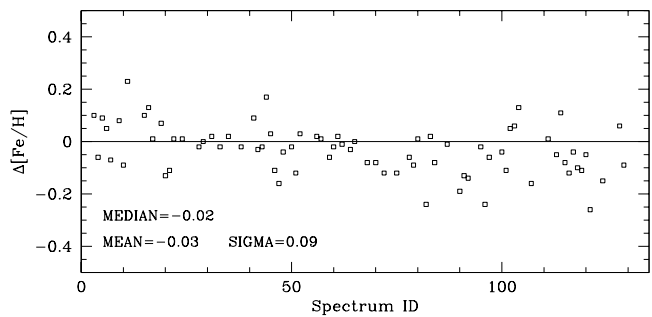,angle=0,width=9cm}
\caption{The difference in [Fe/H] between the two independent measurements
of the two repeated sets of spectra in the $b=-12^\circ$ field.}
\label{repeat}
\end{figure}

Because of a  mistake in the fibre   allocation, in the  $b=-12^\circ$
field a sample  of  $\sim 100$ stars was   observed twice, instead  of
switching to the next target  sample.  This unintentional duplication,
on the other hand, turned out  to be very useful  to estimate the {\it
repeatability} of our measurements.  The  two sets of spectra for  the
same stars  have been reduced in a  fully independent way,  as if they
were different stars,  and  the stellar  parameters were also  derived
independently.  Thus, the differences  in the metallicity obtained for
the same star is not only due to statistical fluctuations, but also to
possible  differences in the  adopted parameters.  Figure~\ref{repeat}
shows the difference in the [Fe/H] of each  star, from the two sets of
observations.  The figure label shows the mean,  median, and spread of
the distribution.  We  checked for correlations of the  $\Delta$[Fe/H]
against any stellar   parameter (T$_{\rm eff}$, $V_{\rm t}$,   [Fe/H],
S/N, ...  )  but to our great surprise,  the only quantity that showed
a mild correlation with   the [Fe/H] difference  is the  spectrum  ID,
meaning the fibre position along  the slit.  Specifically, more than a
{\it trend} what we see is a systematic offset between the first $\sim
65$  stars   ($<\Delta[Fe/H]>\approx     0$)  and   the    next   ones
($<\Delta[Fe/H]>\approx -0.07$).  No physical property  of the star is
associated  with  this   parameter,   and  it   is unlikely that   any
instrumental effect  could explain  this behaviour. The  difference is
instead due  to the  {\it fluctuations}  in the  subjective process of
converging to the best stellar  parameters.  In other words, since the
stellar   parameters,  and  in   particular  the  temperature and  the
microturbulence velocity, produce similar  results on the line-by-line
[Fe/H] abundances, it is possible  to converge on two different  model
atmospheres  (i.e., with  both  different T$_{\rm  eff}$ and different
$V_{\rm  t}$, compensating  each  other)  while  preserving both   the
excitation  equilibrium and  a  constant abundance  with EWs.  The two
models  will give slightly   different mean iron abundance. Therefore,
the resulting  [Fe/H] may differ  by  as much  as $\approx  0.07$ dex,
depending on whether  one starts by iterating  on $T_{\rm  eff}$ until
reaching excitation equilibrium, then fixing the required $V_{\rm t}$,
or one  proceeds in the  opposite direction, first fixing $V_{\rm t}$,
and  then iterating on $T_{\rm eff}$.   According to our records, this
change of procedure occurred  in fact around spectrum  Nr.  65.  While
we  could have re-analyzed the stars  keeping  a uniform procedure, we
preferred to  leave track of the  effect  that such  difference in the
analysis    has    caused    on the       resulting   [Fe/H].   Hence,
$\Delta[Fe/H]<$0.07 dex, is a good  estimate of the mean  fluctuations
due to the subjective part of the analysis.

On the  other  hand, for stars   with  metallicity close to  solar,  a
systematic  error  of $\pm200$  K in   the adopted  T$_{\rm eff}$ (and
corresponding change in the gravity  calculated from eq.~1) implies  a
$\Delta$[Fe/H]=$^{+0.18}_{-0.15}$ dex, for a star  with T=4800 K, and
$\Delta$[Fe/H]=$^{+0.07}_{-0.03}$ dex, for a star  with T=4300 K.   A
systematic error of $\pm0.2$ in the microturbulence velocity implies a
$\Delta$[Fe/H]=$^{-0.12}_{+0.13}$ dex, for both  cool and  warm stars.
A  more  extensive discussion  of systematic   errors in this  kind of
analysis is presented in Lecureur et al. (2008).

\subsection{Stars observed with UVES}

\begin{figure}[hb!]
\psfig{file=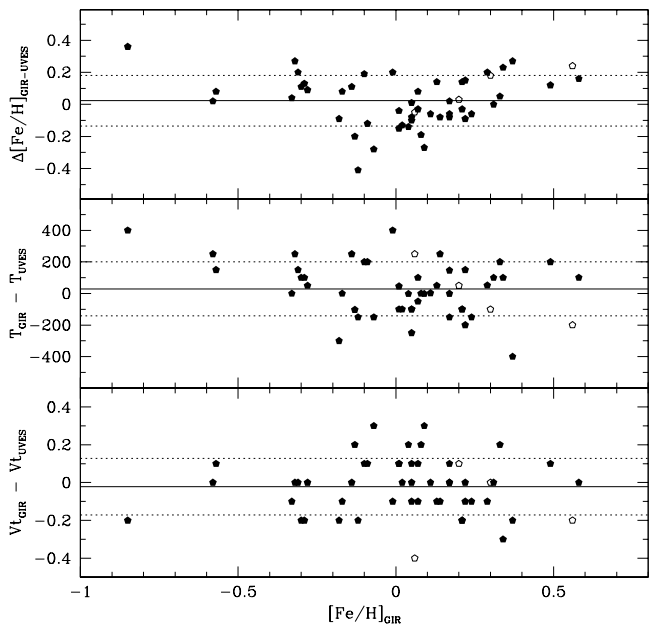,angle=0,width=9cm}
\caption{Comparison between the measured iron abundances, temperature
and microturbulence velocity in the stars  observed both with UVES and
GIRAFFE. Open  symbols    are stars with  larger  dispersion    in the
line-by-line iron  measurements  (mostly metal  rich stars).  The mean
systematic difference (solid line) and the $\pm1 \sigma$ spread around
it (dotted lines) are shown.}
\label{giruves}
\end{figure}

Figure~\ref{giruves} shows,   for the stars    observed also  at  high
resolution with UVES, the comparison between resulting iron abundances
(upper  panel)  and the   most relevant parameters  (middle  and lower
panels). We note  that the stars  observed with UVES  were 58 in total
(c.f.  Table~1 in Lecureur et al. 2007): 13 were Baade's Window clump,
and  another  13 were  giants, 11  more   giants were observed  in the
$b=-6^\circ$  field,  7  in  the $b=-12^\circ$  field   and 14  in the
NGC~6553 field.  However only 50 of  them were discussed in Zoccali et
al.  (2006) and  53 in Lecureur   et al. (2007) because  those studied
focused on  the analysis  of  a few specific lines,  sometimes heavily
blended with telluric lines.

>From the 58 UVES stars, here we exclude  from Fig.~\ref{giruves} the 7
stars  in the $b=-12^\circ$  field because they were never re-observed
with    GIRAFFE, and one  more  clump   star  that  also  failed to be
reobserved with GIRAFFE.  We are thus left with  50 data points. Among
them, open symbols are stars with large dispersion in the line-by-line
iron determination, mostly high metallicity stars, and most likely due
to line crowding.

The systematic offset is  negligible in all  the panels.  The scatter,
again   representative   of  the  statistical  error,   is $\sigma{\rm
[Fe/H]=}\pm 0.16$,  consistent  with our estimates  above. The largest
scatter is in the adopted  excitation temperature, revealing that this
parameter is constrained to no better than $\pm 200$ K.

\subsection{Globular cluster stars}

\begin{figure}[ht]
\psfig{file=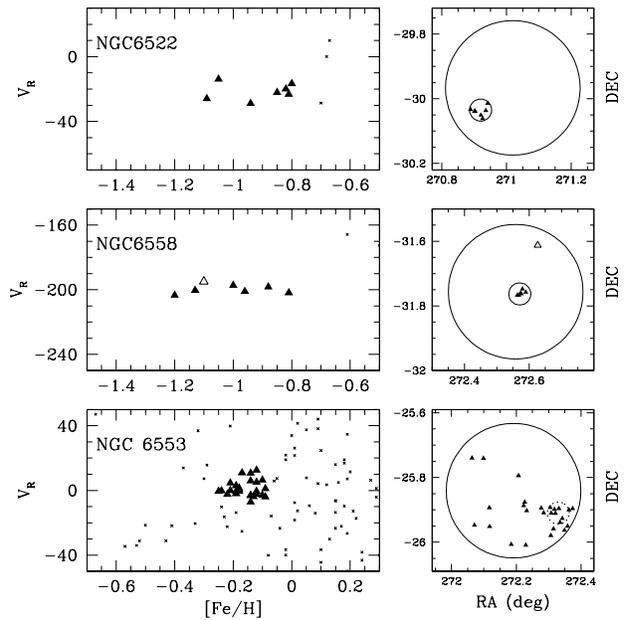,angle=0,width=8.5cm}
\caption{Left panels: iron abundance versus radial velocity for globular 
cluster  stars  included among  our targets  (filled triangles). Bulge
field stars  are also shown as  small  symbols, in order  to emphasize
that while NGC~6522  and NGC~6558 can  be easily separated from  field
stars, some ambiguity is present  in the selection of stars  belonging
to NGC~6553,  due to  its   near solar metallicity  and  low  radial
velocity.  Right panels: position of cluster stars with respect to the
FLAMES field of view (large circle). The small  circle has a radius of
2 arcmin around  the  cluster  center. One star,   shown as   an  open
triangle in  the middle panel,  has   metallicity and radial  velocity
similar  to the other  NGC~6558 members, but  it is very far away from
the cluster center, making it unlikely to be a member. The lower right
panel  shows  once  again that  unambiguous identification  of cluster
members in NGC~6553 is very hard.  }
\label{GCstars}
\end{figure}

Yet another independent test on the internal precision of our analysis
is offered by the stars which are members  of the globular clusters in
these fields.  The left panels  of Figure~\ref{GCstars} show a plot of
radial velocity versus metallicity for all the stars in our fields (in
a  narrow range of   radial  velocity and metallicity)  together  with
globular cluster stars, shown here as  filled triangles.  The location
of cluster stars in the field of view of  FLAMES is shown on the right
side of  the plot.  Cluster  members  were  selected  as target  stars
having [Fe/H] within   $\pm 0.2$ dex  from  the cluster  mean,  radial
velocity within $\pm  10$  km/s from the mean,   and located within  2
arcmin from the   cluster  center.  Baade's Window   contains  7 stars
belonging to the  metal-poor cluster NGC~6522,  and only one member of
NGC~6528, at solar  metallicity and radial velocity  close to 200 km/s
(not shown here).  The  field at $b=-6^\circ$  contains six members of
NGC~6558  (Barbuy et al.   2007).  Finally the NGC~6553 field contains
the eponymous cluster,  but its position  in this plot  falls near the
center  of the distribution  of the field stars, thus  it is harder to
discriminate cluster  from field, and  for this reason the metallicity
spread   of NGC~6553 {\it putative}   members  is not considered here.
Cluster stars should have identical velocity and composition, thus the
observed spread in this plot is a  measure of our (mostly statistical)
error.   For NGC~6522  and  NGC~6558 the 1$\sigma$  spread for cluster
stars is $\sigma [Fe/H]=0.12$ and $\sigma [Fe/H]=0.15$, respectively.

A  complete analysis of  the chemical abundances  of cluster stars has
been presented in   Barbuy et al.  (2007) for  NGC~6558, and it  is in
preparation for NGC~6522.   What we show here is  the iron content  of
cluster stars,  as measured considering  them just  like all the other
field stars (e.g., adopted distance and reddening are  the same as the
mean  ones for the   bulge) and the  details  of the analysis, such as
sigma clipping in Fe lines, etc.,  are suitable to  be extended to all
the  target stars. For this reason,  the actual metallicity of cluster
stars derived here   is not as accurate    as it is  in  the dedicated
papers, though  well within our 1 sigma  error bar.  Cluster stars are
shown here with  the only purpose of helping  estimating our  error on
individual [Fe/H] measurements.

In summary, the three independent estimates  of the internal error via
{\it i)} repeated and  independent analysis, {\it ii)} comparison with
the  UVES results  and  {\it  iii)} globular   cluster stars, indicate
$\sigma [Fe/H]=0.09$, 0.16, and  0.12  dex, respectively.  All   those
estimates include  the smaller ($<$0.06 dex)  statistical error due to
line-to-line  dispersion, but each of them  includes  only a subset of
all   the possible causes  of  errors.  Putting together the different
tests,  and considering that some of  the  systematics (e.g., possible
non LTE effects,   errors in the  model  atmospheres themselves, etc.) 
have not been taken into account here, we can  conclude that $\pm 0.2$
dex is a conservative estimation of our uncertainty on the metallicity
of the individual star,  including both the  effect of statistics  and
systematics.

\section{The distribution functions of the iron abundance}

\begin{figure}[ht]
\psfig{file=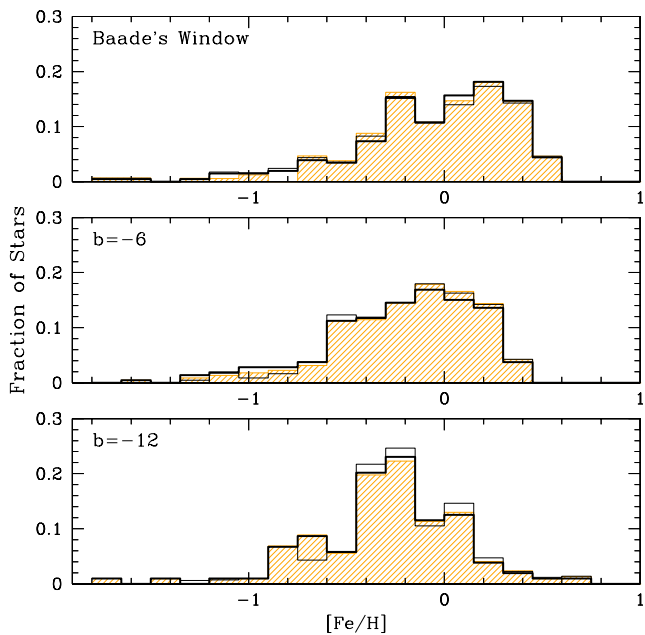,angle=0,width=9cm}
\caption{The raw IDF (thick histogram) compared with the IDF corrected
for color bias, according to method {\it i)} (shaded histogram) and
method {\it ii)} (thin histogram) discussed in Sec.~2.1 .}
\label{mdfcorr}
\end{figure}

The IDFs obtained  in the three fields  along the bulge minor axis are
shown in Fig.~\ref{mdfcorr}, and  the corresponding data are  given in
Table~\ref{allfeh}.  The thick histograms  show the raw IDFs, while the
shaded and the thin open one are the IDF corrected from the color bias
discussed  in  Sec.~2.1,  using  method   {\it   i)} and  {\it   ii)},
respectively. The differences are in fact very small, fully consistent
with our error bars, but we judged important to prove to ourselves that
this kind  of  bias was negligible,  given  the way our  targets  were
selected.  We do not  show here the IDF for  the field around NGC~6553
due to the fact that, as shown in  Fig.~\ref{cmds}, this field has the
strongest differential reddening,  and none of  the reddest stars were
included in our target  list.  Thus, we  believe that, if there is any
bias, in NGC~6553 our sample may be biased against the most metal-rich
stars. In addition, in order to evaluate the fraction of stars sampled
at each color, we had to exclude cluster stars both in the total color
magnitude diagram  and   in the   target sample.   This  task   proved
extremely  hard  in the  NGC~6553   field,  due  to the dimension  and
centrality of the  cluster.  Finally, as shown in  Figs.~\ref{GCstars}
and ~\ref{kinem}, both  the  metallicity and  the radial  velocity  of
cluster stars  sit just in the  middle  of the distributions  of field
stars. For these reasons, we will not include the IDF of this field in
our discussion of the general bulge iron  content.  On the other hand,
the NGC~6553  field,   thanks to its   largest  extinction, will prove
useful in  our analysis of the  disk  contamination (see discussion in
Sec.~8).

\begin{table*}
\caption{Stellar parameters and iron abundance of all the program stars. }
\label{allfeh}
\begin{tabular*}{0.99\textwidth}{@{\extracolsep{0pt}}rrrcccccccrcc}
\hline\hline
\noalign{\smallskip}
QF$^{\mathrm{a}}$ &  ID  & OGLE-ID  &   RA  &   DEC  &  V  &   V-I  &   log g  &   V$_t$  &  T$_{\rm eff}$ &  [Fe/H]  & $\sigma^{\mathrm{b}}$ & cluster? \\
\hline
\multicolumn{13}{c}{Baade's Window}\\
0  &    2   &    423342   &  18:03:50.00  &  -29:55:45.20   &  16.36  &  1.805  &  1.99  &  1.3  &  4650  &   0.46  &  0.38  &  -- \\  
0  &    3   &    423323   &  18:03:48.39  &  -29:56:27.10   &  16.10  &  1.846  &  1.59  &  1.5  &  4200  &  -0.48  &  0.18  &  -- \\
0  &    4   &    412779   &  18:03:43.18  &  -29:59:40.10   &  15.91  &  1.667  &  1.93  &  1.5  &  4850  &  -0.37  &  0.18  &  -- \\
2  &    5   &    412803   &  18:03:46.14  &  -29:58:30.00   &  16.40  &  2.083  &  1.52  &  1.3  &  4000  &   0.51  &  0.34  &  -- \\
0  &    6   &    423359   &  18:03:47.03  &  -29:54:49.20   &  16.17  &  1.768  &  1.92  &  1.4  &  4650  &  -1.23  &  0.23  &  -- \\
... &   ...   &     ...   &  ...          &   ...           &   ...   &  ...    &   ...  &  ...  &  ...   &   ...   &  ...   & ... \\
\hline
\end{tabular*}
\begin{list}{}{}
\item[$^{\mathrm{a}}$] QF is a subjective quality factor, classifying stars into good (0), 
     bad (2) and intermediate (1), according to how unique/degenerate 
     the convergence into the final model atmosphere proceeded. 
\item[$^{\mathrm{b}}$] Line-to-line dispersion around the mean [Fe/H].
\end{list}
\end{table*}

\begin{figure}[ht]
\psfig{file=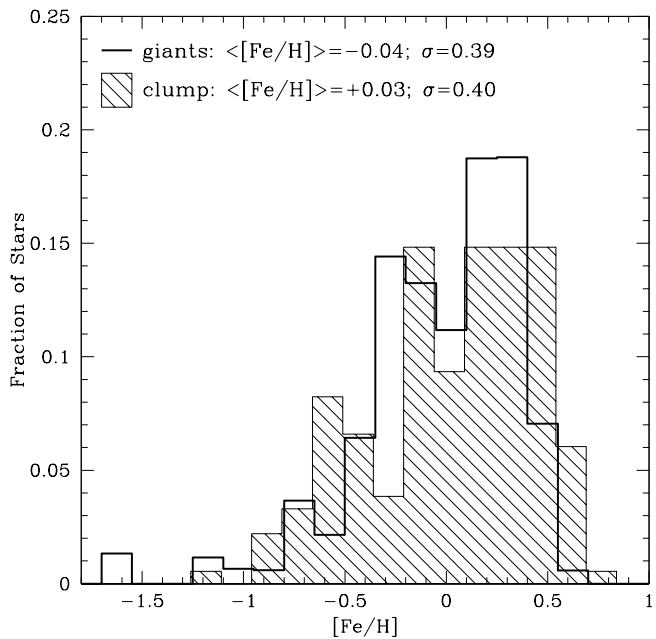,angle=0,width=9cm}
\caption{Comparison between the IDF of Baade's Window as derived from 
giant and red clump stars, the latter from Lecureur et al. (2008).}
\label{mdfbw}
\end{figure}

As mentioned before, for the Baade's Window field two independent (but
homogeneous) sets of  data   are available: the 204  giants  discussed
here, and  another  $\sim$200  red clump   giants observed  within the
guaranteed time reserved to the  FLAMES French consortium. The latter,
extensively discussed  in a companion paper   (Lecureur et al.  2008),
have been reduced   in a very similar   way as the  present data,  and
Fig.~\ref{mdfbw} shows the  comparison  between the  IDFs  of the  two
samples. Although some differences seem to be  present between the two
distributions\footnote{As discussed in  Lecureur  et al.   (2008), the
analysis of the clump stars has been done with an automatic procedure,
based on   the same prescriptions followed   here in a  manual  way. A
discrepancy    as large as  that    shown in Fig.~\ref{repeat}, can be
expected between the two IDFs,  for the same reason,  and it is  still
small when compared to the total uncertainty presented above: in fact,
the means of the two distributions  agree within 0.06 dex.}  Note that
the smaller amount of  metal poor stars in the  clump IDF is expected,
since metal poor stars would not be found in the  red clump but on the
blue side of the  horizontal branch (HB).   However, there are  really
few  metal-poor stars even  in the giant IDF  (only 6 out of 204 stars
have   [Fe/H]$<-1.0$) hence     we    consider   this   bias    rather
negligible. Therefore, in the  following discussion the two  sets will
be combined and  the quoted  Baade's  Window IDF will  result from the
independent analysis of a total of $\sim$400 stars.

\begin{figure}[ht]
\psfig{file=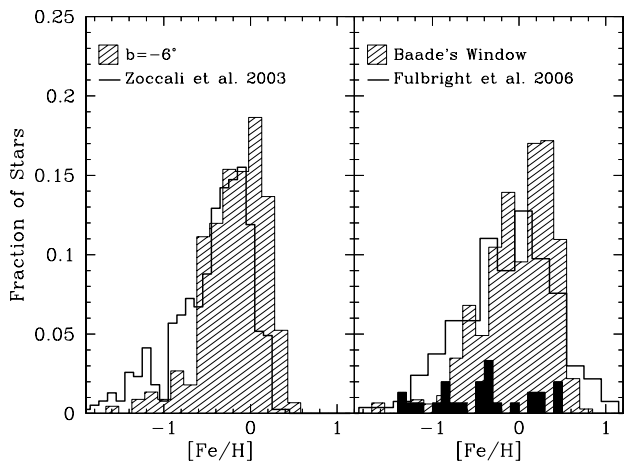,angle=-90,width=9cm}
\caption{The derived IDF is compared with previous measurements, in 
the corresponding  fields.   Left: the   photometric  IDF  by  Zoccali  et
al. (2003)  obtained in the $b=-6^\circ$  field observed here.  Right:
the spectroscopic IDF by Fulbright et al.  (2006) in Baade's Window is
compared  with the  present results.    Also shown as   a black  solid
histogram (arbitrarily normalized) is the IDF of the 27 stars actually
observed by Fulbright et al. (2006) at high spectral resolution. }
\label{mdfcomp}
\end{figure}

Figure~\ref{mdfcomp} shows the   comparison with some of the  previous
determinations of the  IDF of bulge  fields.  The  left panel compares
the  present IDF (shaded) with the  photometric one  by Zoccali et al.
(2003), both relative to  the field at  $b=-6^\circ$. The two IDFs are
different, especially at high metallicity, possibly due to the lack of
calibrating template red  giant  branches  for solar metallicity   and
above.  At the opposite end of the IDF,  the less prominent metal poor
tail   with respect to Zoccali  et  al. (2003)  can  be ascribed to an
innate limit of the photometric method, as  the RGB color becomes less
and less  sensitive to  [Fe/H]  at decreasing metallicity,  hence even
small color errors imply large errors in the derived [Fe/H]. The right
panel  shows  the comparison  with  the spectroscopic IDF  for Baade's
Window  from Fulbright  et     al.   (2006), as  obtained    from  the
recalibration of the Sadler et al.   (1996) IDF.  It  can be seen that
in  both cases the present  spectroscopic IDF  is appreciably narrower
than previous results.  In a sense, this is consistent with our effort
at  reducing  the   errors on     individual measurements.    However,
Fig.~\ref{mdfcomp}  also shows as a  solid histogram the 27 stars that
were   actually  measured   by Fulbright  et    al.    (2006) at  high
spectroscopic  resolution.   Those  are  the stars  that  were used to
recalibrate the Sadler et al.  (1996) IDF obtained from low resolution
spectra.  It can  be appreciated   that  none of   the 27  stars   has
[Fe/H]$>0.5$, despite their selection  of 3 stars  with [Fe/H]$\ge0.5$
in Sadler et al. (1996).  The discrepancy at the metal  rich end is in
a region where the Fulbright  et al.  calibration was  in fact used in
extrapolation.  In  addition, the strong Mg$_2$  features found in the
most metal-rich  and cooler stars  are contaminated by TiO lines (see,
e.g., Fig.~13 by Coelho et al. 2005) and the high end of the Sadler et
al.  (1996) IDF itself probably has  an overestimated high metallicity
tail.

Less obvious is the interpretation  of  the discrepancy at low  [Fe/H]
with  respect to  the  IDF  by  Fulbright et  al.   (2006).  The  high
resolution sample of Fulbright et  al. (2006) contains four stars with
[Fe/H]$<-1$, so  that we know  that  the total sample (88  giants from
Rich 1988) from which those stars were picked (with an on-purpose flat
IDF) had to contain  at least that  number of  stars. This would  mean
that we would expect some $\sim$9 stars in  our RGB sample, whereas we
observe   only 6.   Although  different, the  two   numbers are  still
compatible within the very low statistics considered here.

On the other hand, with some simple calculations we can check that the
number of metal poor stars in the IDF is consistent with the number we
expect from  independent sources.  First,   it is well  known that the
bulge contains RR  Lyrae stars,  classical  tracers of  the metal poor
population.    From the MACHO   (Alcock et  al.    1998)  and OGLE  II
(Collinge et al.  2006) surveys, we know that  there are $\sim 30$  RR
Lyrae per FLAMES field, at $b=-6^\circ$. The total number of red clump
stars in this  field can be estimated  from the CMD in Fig~\ref{cmds}:
there are  4090  stars    within   a box   with  $1.3<(V-I)<2.1$   and
$14.5<I<15.5$. This box includes  both the red   clump and the  RGB at
that  level.  From  the  synthetic  CMD  presented in Zoccali   et al.
(2003; their Fig.~20)  we  know that  only 67\%  of  them, i.e.,  2740
stars, are  actually  red clump stars.  Therefore,   in a FLAMES field
there are 30 RR Lyrae stars for every 2740 red  clump stars, i.e., 1\%
of the total number of stars are expected  to have [Fe/H]$<-1$.  There
could still  be more  metal poor stars   that end up   too blue in the
horizontal  branch  to  pulsate  as RR   Lyrae.  Their  number can  be
estimated from Busso et al.  (2005), who obtained spectra of candidate
extreme   blue HB stars  in the  bulge.  Out of  their  28 targets, 15
(57\%) turned out to be true blue HB stars.  There are 51 extreme blue
HB candidates  in    the CMD   of the  complete   FLAMES field,  hence
$51\times0.57=29$ of them   were   confirmed  spectroscopically.  This
number is  almost identical to the  number of RR  Lyrae, hence another
1\% of the total number of bulge  stars are expected  to be metal poor
enough to end up in the extreme blue HB.

All in all, based on  the known fraction  of stars in the extreme blue
HB and in the RR Lyrae gap, we expect  that at least  2\% of the total
number  of  stars  in   the bulge should   be   metal poor, say   with
[Fe/H]$<-1$. This percentage has to be taken as a lower limit, because
while  we can easily  count RR Lyrae and  extreme HB stars, there is a
narrow range in  color, corresponding to A-type  blue HB stars that is
heavily contaminated by the disk main sequence.

Our IDF at $b=-6^\circ$ is based on $\sim$200 stars and 6 of them have
[Fe/H]$<-1$, fully consistent with the  4 (at least) expected from the
above calculation.  Therefore, even if  the number of metal poor stars
in the IDF presented here might seem very  small, for example compared
with previous  measurements or with  a simple,  closed box  model (see
below), it is consistent with the  number of expected metal poor stars
in the bulge estimated from independent evidence.

In closing   this section  it  is  worth  mentioning  that Johnson  et
al. (2007,  2008) and Cohen et al.   (2008) have recently measured the
chemical abundances of three bulge dwarfs during a microlensing event.
They find metallicities close to [Fe/H]$\sim$+0.5 for all the stars, a
value too high to be consistent with  random extraction of three stars
from our  IDF. This discrepancy is very  puzzling, although it is fair
to mention that several microlensed bulge  dwarfs had been observed by
Cavallo  et al. (2003) finding    metallicities consistent with  ours.
Speculations have been made  that dwarf stars, being unevolved,  might
give the ``true''  IDF, as  opposed  to giants,  whose evolution might
actually depend on their metallicity.  However, at present there is no
indication  that supports such  major  differences in the evolutionary
path of a star at [Fe/H]=-1.0 with respect  to one at [Fe/H]=+0.5.  As
discussed in  Zoccali et  al.  (2003, their  Fig.~13) the  metallicity
dependence of the evolutionary flux along the RGB (i.e., of the number
of stars reaching  the  RGB per  unit  time)  and of the  stellar  RGB
lifetime has opposite  trends, so that  stars of all metallicities are
equally represented along  the RGB.  Cohen  et al. suggest that higher
mass loss in metal rich stars would cause them to leave the RGB before
reaching the level of our samples (at $I\sim  14.5$), then evolving to
the helium white dwarf stage. Were that  true, one would expect a drop
in the RGB  luminosity function which  is not observed (Zoccali et al.
2003, their Fig.21.

We note  that the extremely high  amplification  of these microlensing
events  ($>300$)  indicates that caustic  crossing took  place and the
amplification may not have been uniform over  the stellar surface. The
lens model and the  model atmosphere  should  take these  effects into
account.

\subsection{Disk and Halo contamination in the Bulge fields}

\begin{figure}[ht]
\psfig{file=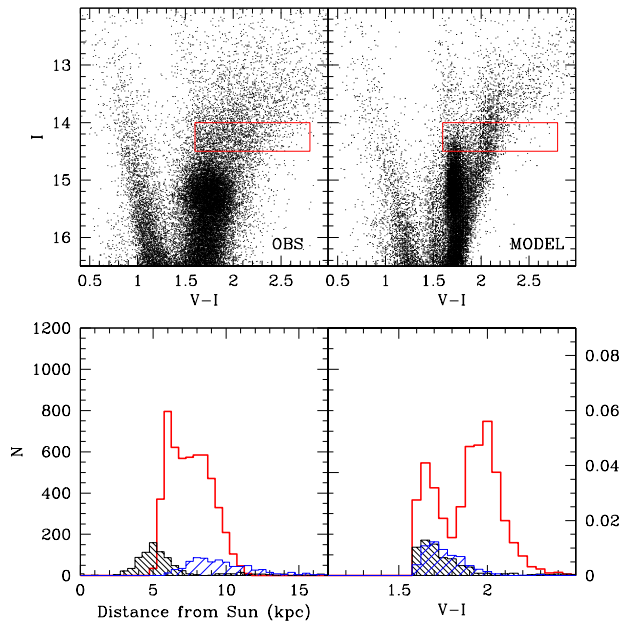,angle=0,width=9cm}
\caption{The upper panels show the observed and model CMD for the
Baade's Window field, together with the box where the targets were 
selected. The bottom panels show the distance and color distribution 
of bulge stars (open histogram) and of thick (light dashed) and thin 
disk (heavy dashed). The disk histograms are scaled to the contamination 
fraction -with respect to the total number of stars- shown in the y-axis 
on the right end side of the plots.}
\label{contbw}
\end{figure}

\begin{figure}[ht]
\psfig{file=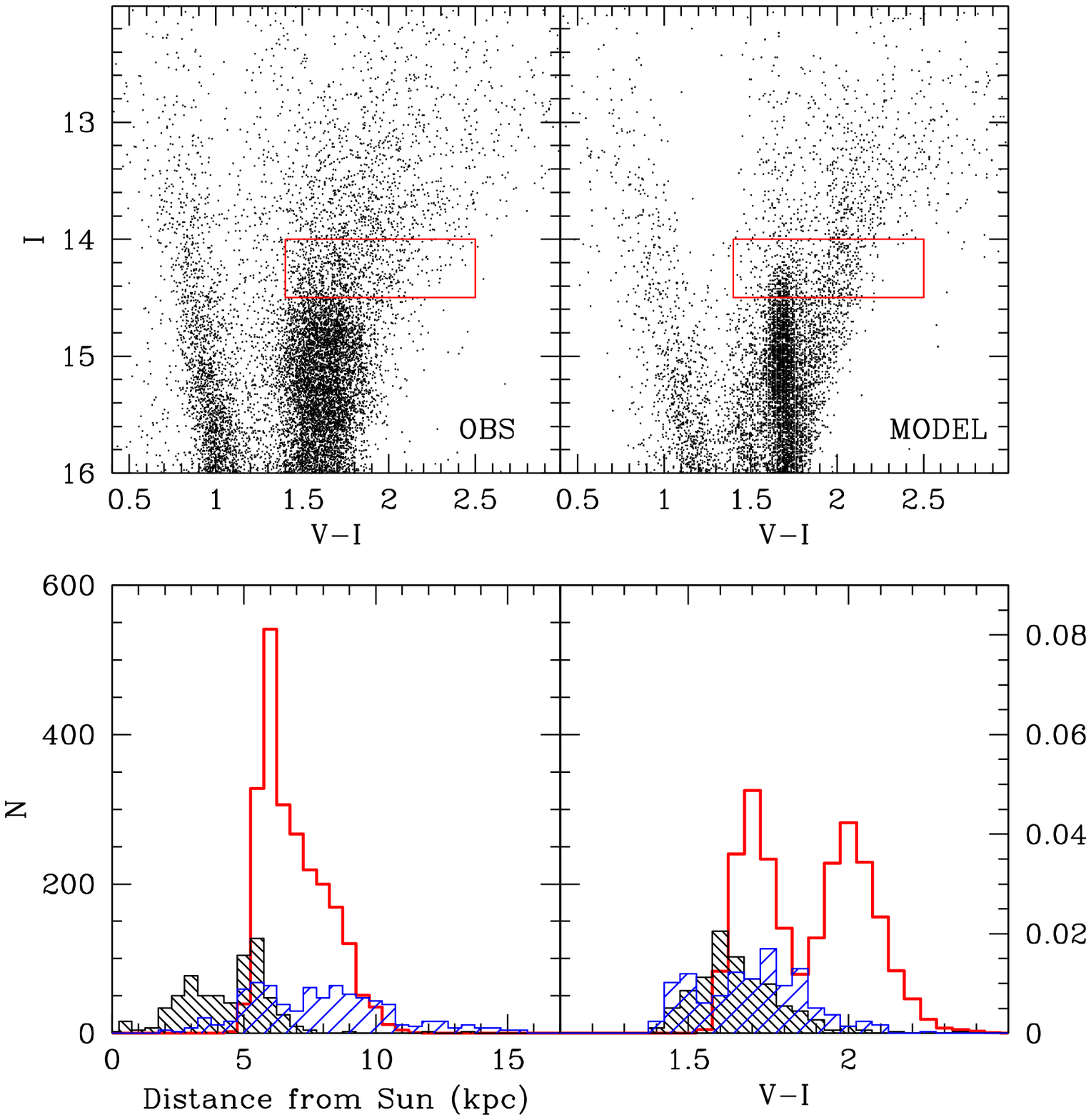,angle=0,width=9cm}
\caption{Same as Fig.~\ref{contbw} for the field at $b=-6^\circ$}
\label{contb6}
\end{figure}

\begin{figure}[ht]
\psfig{file=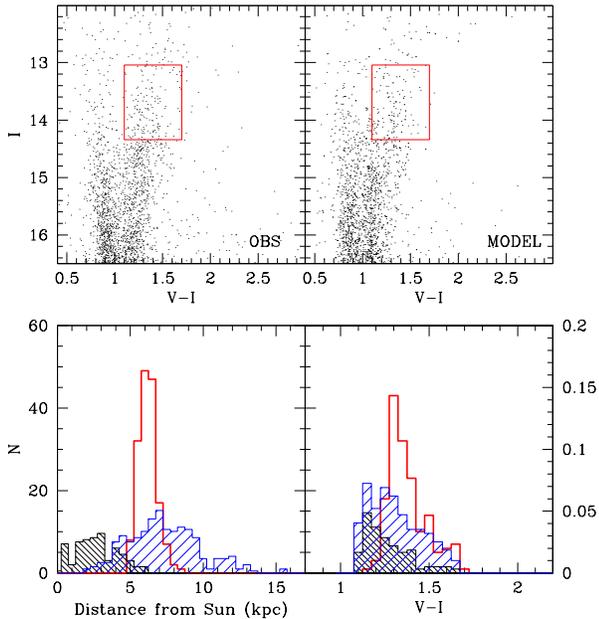,angle=0,width=9cm}
\caption{Same as Fig.~\ref{contbw} for the field at $b=-12^\circ$}
\label{contb12}
\end{figure}

In this section we present our  estimates for the contamination in the
survey  fields coming  from  the thin  and  thick  disk, and from  the
halo. The working  tool for these estimates is  an  updated version of
the  Besan\c{c}on Galaxy model (Robin et  al. 2003) kindly computed by
M.  Schultheis for us.  Simulated  CMDs have  been constructed for the
three fields along the bulge minor  axis.  Small adjustments were made
in the assumed reddening law in order to insure that the simulated red
clump would  coincide in color and   magnitude with the  observed one.
The resulting model CMDs,  together with the  observed ones, are shown
in the   upper   panels  of   Figs.~\ref{contbw},  \ref{contb6}    and
\ref{contb12}.  Clearly, the model CMDs reproduce reasonably well many
characteristics of the observed  CMDs, but significant differences are
also evident.  For example, the giant branches are much broader in the
data   than  in  the   model, possibly because   the    model does not
incorporate small  scale  differential reddening.  Thus, the  relative
contributions  of the various  galactic components to the star samples
in the various fields need to be taken with caution. However, it is 
still the best tool available to analyse the expected contamination of
our sample from (however poorly) known galactic components on the line
of sight.

Stars inside the  observed target box  were selected in the model CMD,
and their distribution in distance, color, and stellar parameters were
analysed.   The lower  panels of  Figs.~\ref{contbw}, \ref{contb6} and
\ref{contb12}  show those  distributions   in distance and  color. The
raw histogram   of  bulge stars is   shown here,  while  the disk star
histograms are scaled  to the fraction  of the total number  of stars,
reported in  the scale on the  right side of the  lower right box.  We
emphasize, then, that  disk and bulge here are  not shown  in the same
scale, in  order to make the  disk  histograms more visible.   The V-I
histogram of bulge stars in the model shows a  clear bimodality due to
the  inclusion of  some clump stars   -those on the  near  side of the
bulge- and a  small number of asymptotic giant  branch stars.  On  the
other  hand   the  data do   not  show  a  bimodality   in  the  color
distribution. The   discrepancy   may  be ascribed  to   the  specific
assumptions in the  Besancon model, such as  the red clump luminosity,
color, and the bulge density law.

Particularly  interesting  is the distance  distribution,  because  it
helps  understanding the evolutionary phase, thus  the gravity, of the
contaminating stars.  One   can see, for   instance,  that in  Baade's
Window the Besan\c{c}on model  predicts contaminating thick disk stars
to be located at the same distance of the bulge.  Therefore, for these
stars the photometric gravity we  assume  in the analysis is  correct,
hence so is the  derived iron abundance. On  the other hand, the  iron
abundance alone cannot  help   us discriminating possible  thick  disk
stars against the bulge ones. It is  also important to remark that, if
the  model  is correct,  and the   thick disk stars  contaminating our
sample are those as far away as the bulge (or even on the other side),
our  present  knowledge   of  the  thick  disk  characteristics  (age,
metallicity, scale height and density) at  that position is very poor.
Actually, the  predicted thick  disk stars within   the bulge are  the
result of the assumption in the Besan\c{c}on model that the thick disk
follows   an exponential  radial distribution,  then   peaking at  the
Galactic center.

Contaminating foreground thin disk   stars are estimated to  be  giant
stars (not dwarfs as one might naively  expect) located mostly between
2 and 5 kpc from the Sun.

The contamination from the halo population turns out to be between 0\%
and 2\%  in all the fields (see Table~\ref{tabcont}), hence it can be
safely neglected.

\begin{table}[h]
\caption{Disk and halo contamination percentage in each field, relative 
to the total number of stars in the target box.}
\label{tabcont}
\begin{tabular*}{0.48\textwidth}{@{\extracolsep{0pt}}lrrr}
\hline\hline
\noalign{\smallskip}
Field                &   Thin Disk   &  Thick Disk   &  Halo  \\
\hline
Baade's Window Clump &  $3.2\pm0.1$  &  $5.8\pm0.1$  &  $0.1\pm0.02$ \\
Baade's Window RGB   &  $6.5\pm0.3$  &  $4.8\pm0.3$  & $<0.1\pm0.06$ \\
$b=-6^\circ$ Field   &  $9.8\pm0.5$  & $11.5\pm0.6$  &  $0.4\pm0.10$ \\
$b=-12^\circ$ Field  & $18.9\pm1.8$  & $59.0\pm2.1$  &  $1.5\pm0.50$ \\
\hline
\end{tabular*}
\end{table}

\section{The IDF as a function of color}

The combination of the metallicity,   kinematic and color  information
for each target  star permits a  better understanding of the behaviour
of the different components of the inner Galaxy.

\begin{figure}[ht]
\psfig{file=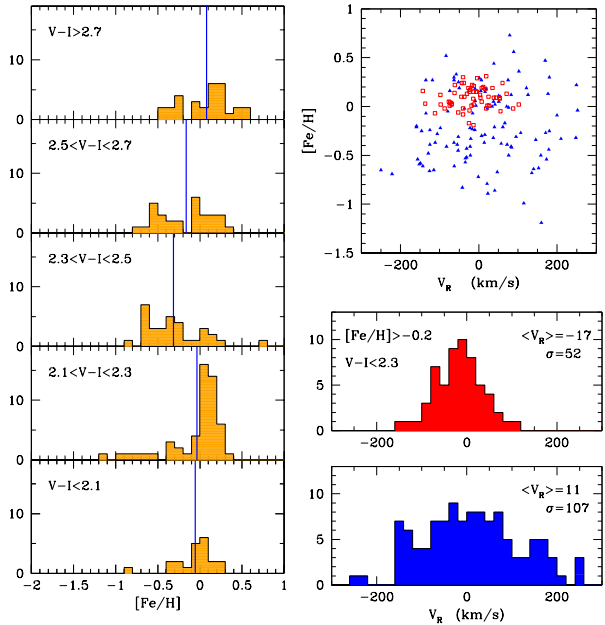,angle=0,width=9cm}
\caption{Left: IDF for the NGC~6553 field as a function of color, from
the bluest stars  at the bottom  to the reddest stars at   the top.  A
vertical line  marks   the mean of   the  distribution.   Upper right:
metallicity versus radial velocity for individual stars. Cluster stars
have been excluded from all these plots.  Empty squares are stars with
V-I$<2.3$,   and [Fe/H]$>-0.2$, while all the   other stars are filled
triangles. Middle right:  radial velocity distribution  for stars with
$V-I<2.3$ and $[Fe/H]>-0.2$ (empty  squares above).  Mean and sigma of
the distribution are shown in the figure label.  Bottom right, same as
above for all  the other field stars  (small  filled triangles  in the
upper right plot).}
\label{colorMDF}
\end{figure}

The  left panels of Fig.~\ref{colorMDF} show  the  IDF of the NGC 6553
field in different   color bins, from blue  (bottom)  to red (top)  as
indicated in the labels. It is well known that,  were the stars all at
the same  distance, i.e. belonging to the  bulge, then more metal rich
giants should  be redder. Therefore,  the IDF should  be progressively
shifted to the  metal rich  side for  increasingly redder  color  bins
(upwards  in  the plots),   with  some possible  spread  introduced by
differential  reddening.  This is   approximately true, except for the
two bluest color bins, that  unexpectedly contain only very metal rich
stars.   If  one looks at the   radial velocity  distribution of those
stars (open squares in the upper right  plot, and middle histogram) it
is clear that they are a colder distribution, with velocity dispersion
of  52 km/s.  On  the contrary, all the  other stars,  shown as filled
triangles in the upper right plot, have a velocity dispersion of 107
km/s.   Note  that suspected clusters stars   are not included   in any of these
plots.  Everything suggests that  the bluest stars in  the CMD  are in
fact contaminating (thin?) disk stars, located on the blue side of the
target  box just because they  are on average  closer to us.  In fact,
there would be   no reason to expect   that the most metal  rich  {\it
bulge} stars should lie  preferentially on the blue  side of the  CMD.
Indeed,  also the Besan\c{c}on model  predicts disk stars to be always
on      the   blue     side         of    our   CMD     target     box
(Figs.~\ref{contbw}-\ref{contb12}).

In this particular field this effect is more evident than in the other
ones because of the larger interstellar extinction  all along the line
of sight, allowing  a color  separation  between bulge  and disk.  The
important conclusion that can be drawn from this  exercise is that the
contaminating (thin?) disk has a very  metal rich IDF, quite different
from that  measured in the   solar neighborhood.   It seems that   the
contaminating disk  is  closer than the bulge  (bluer  in the CMD) but
still quite  far away from  us.  The  existing disk radial metallicity
gradient, then, may explain its higher metallicity with respect to the
solar neighborhood.

\section{A radial metallicity gradient in the bulge}

\begin{figure}[ht]
\psfig{file=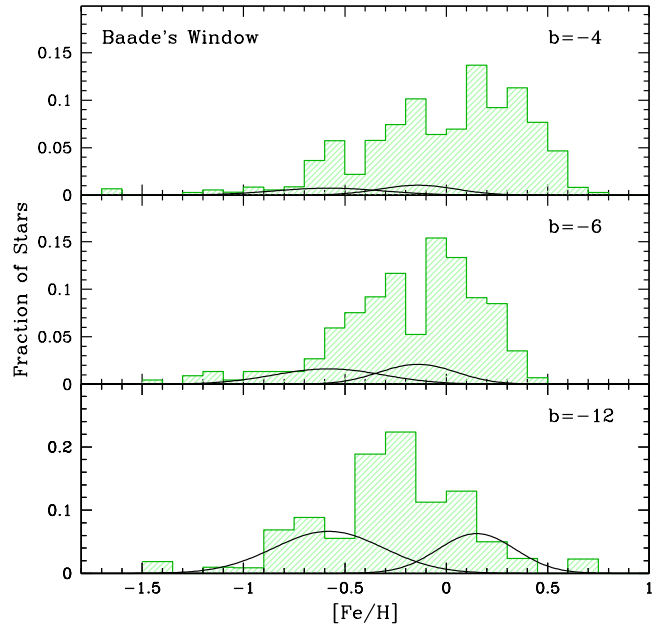,angle=0,width=9cm}
\caption{The obtained IDFs for the three fields along the bulge minor
axis, from the  innermost one (Baade's  Window, top) to the  outermost
one (bottom). The  gaussians show the  IDF of  contaminating thick and
thin disk  stars,  normalized to the expected  contamination fraction,
according  to the      Besan\c{c}on  Galaxy model.  The     thick disk
contamination  percent in the  lower panel has  been reduced at 30$\%$
(as opposed to  the 60$\%$ predicted by the  model) in order  to match
the  number   of  observed  stars with   [Fe/H]$<-0.5$. See   text for
details.}
\label{MDFgauss}
\end{figure}

The final IDFs  for the three fields   along the bulge  minor axis are
shown    in  Fig.~\ref{MDFgauss}.   Overplotted   to   the metallicity
distribution   of  bulge   stars   (histograms)  are  two    gaussians
qualitatively showing the  estimated  contamination by thick  and thin
disk.    The  gaussians  have  indeed   the   mean  and sigma   values
characteristics of   the thick (Reddy   et  al.  2006) and   thin disk
(Nordstr\"{o}m  et al.   2004) IDF,  in the  solar neighboorhood.   As
discussed  above, very likely  the contaminating disk stars are closer
to the bulge than to us, but the disk radial gradient for giant stars,
i.e., intermediate-age and old  disk,  has never been measured,  hence
where exactly these gaussians would lie is not very well known.

\paragraph{Baade's Window}. The IDF for this field has been derived from 
the  combination of both clump   and giant stars ($\sim$400).  Despite
the  uncertainty on the  mean   metallicity of the  contaminating disk
stars, it is clear that their number is negligible in this field.
 
\paragraph{Field at $b$=--6$^\circ$}. The IDF  for this field has been
derived from 213 giant  stars.  Again, the relative disk contamination
is low in this field, and would  not have a  significant impact on the
shape  of the derived bulge IDF.    The comparison with Baade's Window
reveals a difference in the  mean metallicity, suggestive of a  radial
metallicity  gradient,   with the    IDF    mean value   going    from
$<$[Fe/H]$>=+0.03$     at   $b=-4^\circ$    to   $<$[Fe/H]$>=-0.12$ at
$b=-6^\circ$.   More specifically, it  would  seem that rather  than a
solid shift towards more metal poor mean values,  it is the metal rich
stars that gradually disappear, while  the metal poor ones are  always
roughly in the same position.  On the same line,  it is interesting to
note that there is some indication of a bimodality in  the IDF of this
field.

\paragraph{Field at $b$=--12$^\circ$}.   The interpretation of the IDF
for this field, resulting from the observations of 104 stars, is a lot
more  complicated due to  the highest  fraction  of contaminating disk
stars.  In this case it is more  important to establish what should be
the mean metallicity of the contaminating stars.  Regarding thick disk
stars, the Besan\c{c}on model predicts them to make up about 60\% of the
observed stars.  However, if  thick disk  stars  are as metal poor  as
they are seen in the solar neighboorhood, they cannot be as many, just
because we do not  see as many metal poor  stars at all.  If the inner
thick  disk is as metal  poor as it is in  the  solar neighborhood, it
cannot account for more than 30\% of the  total number of stars.  This
(30\%)  would be  the   metal   poor gaussian   shown  in the    lower
panel. Alternatively, thick disk stars are a  bit more metal rich than
in the  solar neighborhood,  probably because they   are closer to the
center (Fig.~\ref{contb12}).  Then in this case they  could be as many
as 60\%, perhaps.   Regarding thin  disk stars,  they are expected  to
make the 20\% of the total  number of stars.  In this case we are more
inclined to  think that they should lie  at the metal  rich end of the
distribution,      because: {\it  i)}       it  will   be   shown   in
Fig.~\ref{colorMDF} that the contaminating thin disk seems to be indeed
very  metal  rich; and {\it  ii)}  because Fig.~\ref{kinem} shows that
there is  a very cold component  at the metal  rich end of this field.
All in all, while it is impossible to conclude what the true bulge IDF
is in this  field, we  can conclude that   the presence of the  radial
gradient seems confirmed in this field.  Indeed,  if thick disk stars are
as metal poor we as we have drawn them in  the figure, then the  mean bulge
IDF is $<$[Fe/H]$>=-0.26$, lower than in the  innermost fields.  Even if
thick disk stars are more metal rich, regardless of how many they are,
then  the mean metallicity  of the remaining  bulge stars  can only be
even lower.   \\

The  discussion  above  draws  our attention to    the  fact that  our
knowledge  of the  disk properties, far   away from the  Sun, is still
extremely  poor.  The Besan\c{c}on  model predicts  a  large amount of
thick disk  stars in the central  region of our Galaxy. However, there
is certainly a hole in the  HI and CO  distribution inside $\sim$3 kpc
(e.g., Dame et al 2001), and we know that in most barred galaxies disk
stars  are  {\it cleaned  up}  in   the central  region.  The
Besan\c{c}on  model does  include a  central  hole  in  the thin  disk
distribution,  but its thick   disk  has  a pure  exponential   radial
distribution.   Does the real thick disk  follow the thin disk and gas
distribution, or does it keep growing toward the center?
 
On one hand,    this  emphasizes the  importance  of   gathering  more
information about the inner disk, in order  to understand not only the
properties  of the disk itself but  also of  other galactic components
affected by disk contamination. On the other  hand, the lower panel of
Fig.~\ref{MDFgauss}, if hard to interpret in terms of bulge IDF, poses
already   important  constraints   on  the  properties  of  the  inner
disk.  Namely,  if   as  much as   60\%  of   the   observed stars  at
$b=-12^\circ$ belong  to the thick  disk, then  their metallicity {\it
must} be definitely  higher than it is in  the solar neighborhood, and
possibly also much narrower.

Finally, we note that while we found indications  of a radial gradient
between $b=-4^\circ$  and $b=-12^\circ$, the  results by Rich, Origlia
and Valenti (2007) indicate  a flattening between ($l,b$)=($1,-4$) and
($l,b$)=($0,-1$).  A flattening of  the radial  gradient in the  inner
bulge below  $b=-4^\circ$ was also   obtained  by Ram\'{\i}rez et  al.
(2000) from low resolution spectroscopy of  giant stars. Also Tyson \&
Rich   (1993) using Washington   photometry   found a radial  gradient
outside $b=-6^\circ$  and a flattening (or a  slight  turnover) in the
inner part.

\section{Metallicity versus kinematics}

\begin{figure}[ht]
\psfig{file=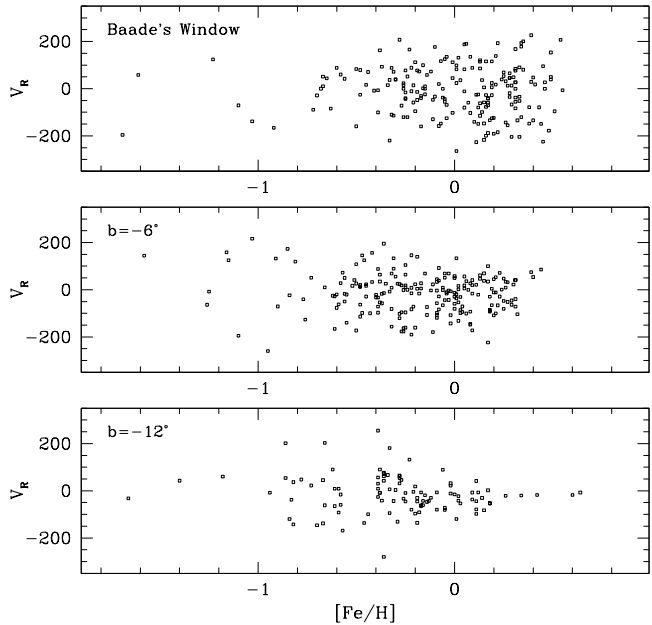,angle=0,width=9cm}
\caption{Metallicity versus radial velocity for individual stars
in the three bulge fields along the minor axis. Globular cluster stars
shown in Fig.~\ref{GCstars} have been removed from this plot. }
\label{kinem}
\end{figure}

Figure~\ref{kinem} shows the radial velocities versus metallicity for
bulge field stars in the three fields. A couple of important pieces of 
information can be extracted from such a plot. 

First, as expected, the velocity  dispersion goes down along the bulge
minor axis,   being $\sigma_{\rm  RV}={105}$  km/s in  Baade's Window,
$\sigma_{\rm RV}=84$ km/s in  the $b=-6^\circ$ field, and $\sigma_{\rm
RV}=80$ km/s   in the  field  at  $b=-12^\circ$. The  latter would  be
further  reduced to $\sigma_{\rm  RV}=60$  km/s  if the  5 stars  with
absolute velocity $|V_{\rm RV}|>150$ km/s are rejected (e.g., if they
were halo stars).

Second, the  velocity dispersion of  the metal rich  tail is extremely
different in the three fields along  the minor axis, being hotter than
the  metal-poor one  in the  innermost field, about   the same  in the
intermediate   one,  and extremely cold  in   the outermost field. The
latter field being heavily contaminated by disk stars, we are inclined
to think that the metal  rich tail is in fact  made by thin disk stars
(see discussion at the end of Sec.~7). On the  contrary, since the two
innermost  fields  both  have  negligible  disk   contamination,   the
interpretation of such a different kinematical  behaviour of the metal
rich  component  with respect to the   metal poor one  is   not at all
straightforward. A detailed analysis  of the bulge kinematics from the
present data  will  be  presented in   Babusiaux et  al.  (2008,  {\it in
preparation}).

\section{Discussion and Conclusions}

\begin{figure}[ht]
\psfig{file=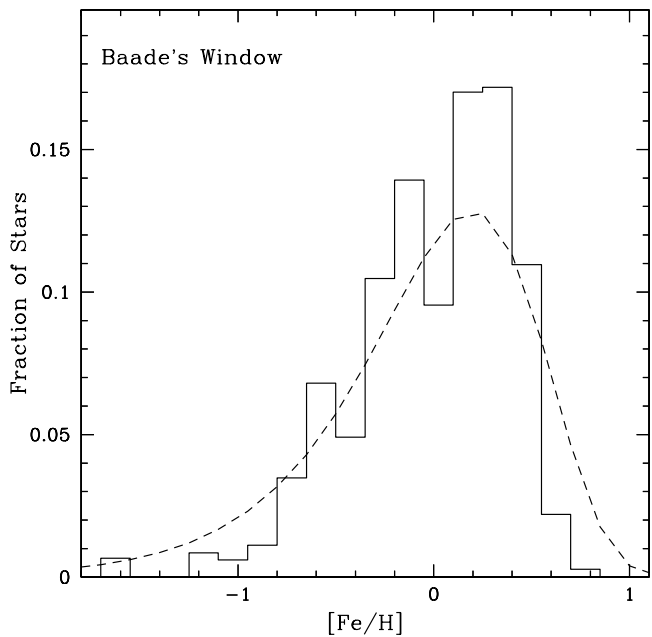,angle=0,width=9cm}
\caption{The observed bulge IDF in Baade's Window compared with a simple, 
one-zone model  with   an   assumed iron yield of $y_{\rm Fe}$=+0.03.}
\label{closedbox}
\end{figure}

Figure~\ref{closedbox} shows  the  [Fe/H] distribution in  the Baade's
Window field with superimposed the  distribution function of a simple,
one-zone, closed-box model of chemical  evolution with an assumed iron
yield $y_{\rm Fe}$=+0.03.  The simple model  has been normalized to 1,
plotting  the fraction    of  the  total   number of   stars at   each
metallicity,  as for  the   observed distribution.  Rich (1990)  first
noted that such simple chemical evolution model is a fairly good match
to the bulge data,  in his case  the  Rich (1988)  data.  As shown  in
Fig.~\ref{closedbox}, this is still the case for the data presented in
this paper. However, at a closer look the match does not look perfect:
the observed distribution  appears to be a   little narrower than  the
model one,  and would be even  narrower after deconvolving  it for the
observational errors.  Moreover,  the  observed distribution  shows  a
sharper cutoff at  high   metallicity, compared  with  the  closed box
model. A small deficit at the  low-metallicity end with respect to the
model was noticed by Zoccali et al.  (2003) for their photometric IDF.
A small deficit was also found by Fulbright,  McWilliam \& Rich (2006)
for  the   original Rich  (1988) sample  of    bulge K  giants, having
recalibrated the old data using Keck/HIRES high resolution spectra for
a subsample of the stars.  But overall, they find quite good agreement
with a  closed-box  model, once either the   mean  or the  median iron
abundance is used for the yield.

Undoubtedly, a closed-box  model  provides a rather  satisfactory {\it
qualitative} match  to the data.   Should  we conclude that  the bulge
really evolved as a closed box?   Certainly not.  In  such a model the
bulge starts its evolution  with its whole  mass in gaseous form,  and
proceeds with star formation till gas exhaustion. Thus, the closed-box
model describes the chemical evolution of  the classic {\it monolithic
collapse}  model.  In modern  scenarios for  the bulge formation,  via
either  merging  of smaller entities  or  via  disk instabilities, the
bulge is assembled  gradually, rather than being  already in one piece
from the beginning. Thus,  the bulge ``box''  was open with respect to
gas (and stars) accretion.  Moreover, most likely  it was also open in
the   opposite  direction,   i.e.,   ejecting  gas   and   metals  via
supernova/AGN driven winds. We expand here on this latter aspect.

The iron yield from theoretical  stellar nucleosynthesis is subject to
large uncertainties, which are difficult  to reduce without help  from
observations.  The iron  yield from individual massive stars exploding
as  Type  II supernovae  is   critically dependent  upon  the  precise
location of the {\it mass cut} between the compact remnant and the
supernova ejecta, which in  fact cannot be  reliably predicted. In the
case of Type Ia supernovae (SNIa), it is their total number (and their
distribution of delay times) as a result of  turning a given amount of
gas into stars that can hardly be predicted only from first principles
(e.g., Greggio 2005).  For these reasons, an empirical estimate of the
iron  yield  may be especially  helpful.   Such an opportunity  can be
exploited in the case  of clusters of  galaxies, which indeed are more
likely  to have  retained all  the   stars, gas and  metals that  have
participated in the evolution. Thus, combining the iron content of the
intracluster medium from X-ray observations,  with that in stars  from
optical  observations of   cluster galaxies, one   finds that clusters
contain $\sim    0.015\;M_\odot$ of  iron  for each     $B$-band solar
luminosity of the cluster galaxies  (Renzini 1997; 2004). We shall now
explore  the consequences of assuming  that  this empirical iron yield
applies also to the Milky Way bulge.

Most of the iron in clusters of  galaxies was produced by the dominant
stellar population, i.e. by stars  in early-type galaxies that  formed
at  $z\gsim 2$ (for a  review,  see Renzini 2006),  i.e., by  galactic
spheroids.  With an age of  over $\sim 10$ Gyr  (Zoccali et al. 2003),
also the stars in the bulge ``formed at $z\gsim 2$'', and the bulge is
a spheroid. Thus, our assumption of a similar  iron yield in the bulge
as in  clusters  is quite reasonable.   Now, with   a present $B$-band
luminosity of $\sim 6\times   10^9\;L_{\rm B,\odot}$ (Kent,  Dame,  \&
Fazio 1991), the  bulge stellar population  should have produced $\sim
6\times 10^9\times 0.015 = 9\times 10^7\;M_\odot$ of  iron. But with a
mass of $1.6 \times 10^{10} M_\odot$  (e.g., Han \& Gould 1995, Bissantz 
\& Gerhard 2002, Sumi et al. 2006) and a mean iron abundance (in  mass)
$<Z_{\rm Fe}>= 0.0018$ (as the average of the individual 
$Z_{\rm Fe} = Z_{\rm Fe,\odot} \times 10^{\rm [Fe/H]}$ in Baade's
Window) the bulge  contains   today only  $\sim
2.9\times 10^7\; M_\odot$ of iron, i.e., about a factor of 3 less than
it should have produced.  Therefore,  under this assumption the  bulge
would  have ejected $\sim 70\%$ of  the iron  it had produced (Renzini
2004).

Chemical   evolution models for the   bulge  that relax the closed-box
approximation   not only with regard  to  bulge assembly  (as in e.g.,
Matteucci, Romano \& Molaro  1999), but also  allowing for bulge winds
are now appearing in the literature (Ferreras,    Wyse, \& Silk, 2003;
Ballero et  al.  2007;  Tsujimoto 2007).  The  bulge IDF  predicted by
Ballero et al.  (2007) qualitatively agrees with the one measured here
(c.f., their Fig.~3).  However, these    models involve several   free
parameters, which are  needed to describe the  rate  at which  new gas
(and stars) are added  to the growing bulge,  the star  formation law,
the IMF, the stellar  nucleosynthesis, the distribution of delay times
for SNIa's, and the  onset and strengths of  the winds.  Some of these
parameters  produce  similar changes on   the predicted IDF making the
comparison between observed and model  IDF not sufficient to constrain
the whole formation scenario.  In  addition, the models predict a {\it
global} IDF  for the  whole bulge.  Due  to the  presence of a  radial
metallicity  gradient, a direct   comparison with observations  is not
straightforward.   In  view  of    these  difficulties,  it  is  worth
summarizing here what  are the major, purely observational constraints
on the formation and evolution of the Galactic bulge.

Zoccali et al. (2003) have  shown that a simulated CMD  with an age of
13 Gyr  (that includes  the  bulge metallicity distribution)   gives a
fairly good  match to the  bulge  CMD. In  particular, this good match
includes  the luminosity difference  between the horizontal branch and
the main sequence turnoff, a classical  age indicator. However, due to
metallicity,  reddening,  and distance  dispersion, the  bulge turnoff
cannot be located to better than 0.2-0.3  mag, corresponding to an age
uncertainty of $\sim 2-3$ Gyr.  Conservatively, we take the age of the
bulk  of bulge stars to  be  in excess of 10  Gyr,   and even so  this
implies that star formation and chemical enrichment had to be confined
within a time interval definitely shorter than the age of the universe
at a lookback time of  10  Gyr, or  $\lsim 3.7$  Gyr according to  the
current  concordance cosmology. If  the bulk of  bulge stars formed in
the cosmic time interval  corresponding to redshift  between 3 and  2,
then star formation  cannot have taken  much more  than  $\sim 1$ Gyr.
Thus, the  main uncertainty    affecting the  duration of   the   star
formation in the   bulge comes from  the  uncertainty in  its age: the
older the age, the shorter the star-formation era.

The second constraint  on the formation  timescale of the bulge  comes
from  the  observed $\alpha$-element enhancement  (McWilliam  \&  Rich
1994,  2003; Barbuy  et  al. 2006; Zoccali  et  al. 2004, 2006; Lecureur  et
al. 2007; Fulbright et al. 2007), once this is interpreted as a result
of the interplay of   the fast delivery of   iron-poor nucleosynthesis
products   of  massive stars by   SNII's,  with the   slow delivery of
iron-rich products by SNIa's.  Again,   a star formation timescale  of 
approximately 1 Gyr is  generally derived from chemical  evolution models,
which typically assume a distribution of SNIa delay times from Greggio
\& Renzini (1983). Thus, the derived timescale is modulo the adopted 
distribution   of   SNIa    delay  times.    Other equally   plausible
distributions (e.g. Greggio 2005)  would have given shorter  or longer
timescales. Thus, until  the actual mix of  SNIa progenitors  is fully
identified, we shall remain with this  uncertainty on how to translate
an $\alpha$-element overabundance   into a star  formation  timescale.
All in  all, combining the  age  and the $\alpha$-element  enhancement
constraints, it is  fair to conclude  that the formation of  the bulge
cannot have taken much  more that $\sim 1$  Gyr, and possibly somewhat
less than that.

In addition, the indications  of  a radial metallicity  gradient found
here would  argue against the formation via   secular evolution of the
bar, because obviously the vertical heating that transforms a bar into a
{\it pseudo}-bulge  would not act preferentially  on metal poor stars.
However, combining our result with  previous ones on the inner  bulge,
at the moment  there is  evidence of  a  flat metallicity distribution
inside $\sim$ 600  pc, and a  radial  gradient outside.  Should  those
findings   be confirmed, they     might indicate  the  presence  of  a
double-component bulge, an inner   {\it  pseudo}-bulge, and an   outer
classical one,  as already found by  Peletier et al. (2007) within the
SAURON survey of galaxy bulges.


Finally, concerning the bulge chemical evolution,  from the IDF we can
certainly conclude that the bulge  must have accreted primordial  gas,
due to the lack of metal  poor stars with  respect to the simple model
prediction  (the so-called G dwarf problem,  solved with the inclusion
of  some  infall in  the model) and  must  have  ejected a substantial
fraction of  the iron  it produced (outflow).    In addition, from the
overabundance of $\alpha$-elements quoted above  we can conclude  that
it  cannot have accreted  stars already significantly enriched by SNIa
products, such as disk stars, or stars born in small galactic entities
similar to the surviving satellite galaxies  in the Local Group (e.g.,
Venn et al. 2004).

\begin{acknowledgements}
We thank Yazan Momany for providing the astrometric and photometric 
catalogue for the NGC~6553 field.

We thank Mathias Schultheis for providing us the results of an updated
version of  the Besan\c{c}on Galaxy model.   This work has been partly
funded by the FONDAP Center for Astrophysics 15010003  (MZ and DM) and
by Projecto FONDECYT Regular \#1085278.  DM  and BB acknowledge the
European Commission's ALFA-II   programme, through its funding  of the
Latin-american  European  Network   for   Astrophysics  and  Cosmology
(LENAC).   BB  acknowledges   grants  from   CNPq  and   Fapesp.    SO
acknowledges the Italian  Ministero dell'Universit\`a  e della Ricerca
Scientifica e Tecnologica (MURST)  under the program 'Fasi iniziali di
evoluzione dell'alone e del bulge Galattico' (Italy).
\end{acknowledgements}






\end{document}